\documentclass[fleqn,twoside]{article}

\usepackage{espcrc2mm}
\usepackage{graphicx}
\usepackage{amsmath}
\usepackage{amssymb}

\title{Transition from direct to sequential two-proton decay in $s$-$d$ shell
nuclei}

\author{%
T.A.~Golubkova\address[sunc]{Advanced Educational and Scientific Center, Moscow State University, Kremenchugskaya 11, 121357 Moscow, Russia},
X.~Xu\address[unig]{Justus-Liebig University, Giessen, Germany}$^,$%
\address[gsi]{GSI Helmholtzzentrum  f\"{u}r Schwerionenforschung, 64291 Darmstadt, Germany}$^,$%
\address[bei]{School of Physics and Nuclear Energy Engineering, Beihang University, 100191 Beijing, China},
L.V.~Grigorenko\address[dub]{Flerov Laboratory of Nuclear Reactions, JINR,  141980 Dubna, Russia}$^,$%
\address[mifi]{National Research Nuclear University ``MEPhI'', Kashirskoye shosse 31, 115409 Moscow, Russia}$^,$%
\address[kur]{National Research Centre ``Kurchatov Institute'', Kurchatov sq.\ 1, 123182 Moscow, Russia},
I.G.~Mukha\addressmark[gsi]$^,$\addressmark[kur],
C.~Scheidenberger\addressmark[gsi],
M.V.~Zhukov\address[chal]{Department of Physics, Chalmers University of Technology, S-41296 G\"{o}teborg, Sweden}
}

\begin{document}

\begin{abstract}
Transitions among different mechanisms of two-proton decay are studied in general. The introduced improved direct-decay model generalizes the  semi-analytical models used before and provides flawless phenomenological description of three-body correlations in $2p$ decays. This is demonstrated by examples of the low-lying $^{16}$Ne state decays.
Different forms of transition dynamic are shown to be highly probable beyond the proton dripline for the $s$-$d$ shell nuclei. It is demonstrated that transition dynamic of $2p$ emitters can provide means for extraction of a width of the ground-state resonance of a core+$p$ subsystem of the core+$2p$ system. Practical applicability of the method is demonstrated by properties of the $^{14}$F ground state  derived from the $^{15}\mbox{Ne}\rightarrow ^{\,13\!\!}\mbox{O}+2p$ decay data and of the $^{29}$Cl ground state derived from the $^{30}\mbox{Ar}\rightarrow ^{\,28\!\!}\mbox{S}+2p$ decay data.

\vspace{3mm}


\vspace{1.5mm}


\vspace{1.5mm}


\vspace{1.5mm}

\noindent \textit{Date:} \today.

\end{abstract}

\maketitle


\section{Introduction}

\begin{figure}[tb]
\begin{center}
\includegraphics[width=0.44\textwidth]{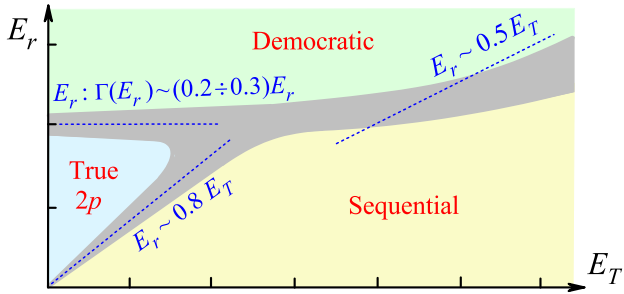}
\end{center}
\caption{(Color online) Sketch of regions with different $2p$ decay regimes (the true, democratic, and sequential mechanisms are indicated by different colors) in the  $\{E_T,E_r\}$ dimensions. The transition region separating these regimes is shown by gray color. The dotted lines forming the ``skeleton'' of the transition region correspond to the displayed ratios between the $2p$-decay energies $E_T$ in a three-body system and $p$-decay energies $E_r$ of the g.s.\  of the core-$p$ subsystem. }
\label{fig:phase-diagram}
\end{figure}

In the recent years there were important advances in the studies of two-proton ($2p$) radioactivity \cite{Goldansky:1960}, and in general of \emph{true two-proton decays} \cite{Pfutzner:2012}. These studies belong to a general trend of spreading our knowledge about limits of existence of nuclear structure as far beyond the driplines as possible. The true $2p$ decay mechanism is realized under certain conditions, in particular when the relation  between one- and two-proton separation energies  ($S_p$ and $S_{2p}$, respectively) makes \emph{sequential decay} not possible: $S^{(A)}_{2p} < 0$ and $S^{(A)}_{p} > 0$ (for nuclei consisting of $A$ nucleons). In the light proton-rich nuclei, the true $2p$ decay can also be described by the mechanism of \emph{democratic decay} \cite{Bochkarev:1989}, which results in distinct correlations between the decay products and different systematics of lifetimes compared to $2p$ radioactive decays \cite{Pfutzner:2012}. The true $2p$ and democratic decays are often characterized as \emph{direct} decays to continuum in contrast with \emph{sequential} decay mechanism, which requires population of intermediate narrow states. Transition to the democratic decay dynamic from the true $2p$ mechanism is typically defined by the width $\Gamma_r$ of the resonance ground  state with the decay energy $E_r$ in the core+$p$ sub-system of the three-body system  core+$p$+$p$. So, in the most common case the dynamic of the $2p$ emission is fully characterized by the three parameters: (i) the $2p$ decay energy $E_T=-S^{(A)}_{2p}$, (ii) the ground state (g.s.) decay energy in the core+$p$ subsystem $E_r=-S^{(A-1)}_{p}=S^{(A)}_{p}-S^{(A)}_{2p}$, and (iii) the width $\Gamma_r$ of the intermediate state. As the width $\Gamma_r$ is a function of $E_r$, then the delimitation of different forms of the $2p$ decay dynamic for the selected nuclide can be depicted by a kind of ``phase diagram'' in the $\{E_T,E_r\}$ space, see Fig.\ \ref{fig:phase-diagram}.

The transitions between different forms of dynamic were fragmentarily considered in the previous works \cite{Grigorenko:2003b,Mukha:2012,Pfutzner:2012}. In the present work we consider systematics of the transition types ``True $2p$''$\leftrightarrow$``Democratic $2p$'' and ``True $2p$''$\leftrightarrow$``Sequential $2p$''. We demonstrate  that these transitions could occur regularly in decays of g.s.\  of $s$-$d$ shell nuclei beyond the proton dripline. The specific motivation of the present work is quite pragmatic. Any phenomenon of the transition type demonstrates abrupt changes of its properties in response to minor variation of the parameters. Such a sensitivity can be used for more precise derivation of the nuclear parameters or for imposing limitations on their relations. Among the parameters $\{E_T,E_r,\Gamma_r\}$, the most interesting from practical point of view is the sensitivity to the value of $\Gamma_r$.

The measurements of resonance positions can be performed with a good precision by using the data obtained with reasonable resolution and statistics.
In contrast, the experimental derivation of the state widths with comparable precision requires much better resolution or/and statistics. This could be not practically attainable for exotic dripline systems. Moreover, there exists a broad ``blind spot'' in the experimental derivation of the g.s.\ resonance widths  in the interval from $\sim 1 \mbox{ ps}^{-1}$ to $\sim 10$ keV. The widths smaller than $\sim 10$ keV do not seem practically accessible by modern invariant/missing mass experiments, while other methods (in-flight, plunger, \emph{etc.}) are available only for long-living systems with widths $\lesssim 1 \mbox{ ps}^{-1}$. We suggest an indirect method for obtaining information on the width $\Gamma_r$ of the proton subsystems in the case of transition dynamic of two-proton emission. The practical feasibility of the method is demonstrated on the examples of $^{14}$F g.s.\ populated in the $2p$ decay of the $^{15}$Ne g.s.\ \cite{Wamers:2014} and of $^{29}$Cl g.s.\ populated in the $2p$ decay of $^{30}$Ar  \cite{Mukha:2015}.


\section{Dynamic of three-body systems near the proton dripline}

\begin{figure}[tb]
\begin{center}
\includegraphics[width=0.49\textwidth]{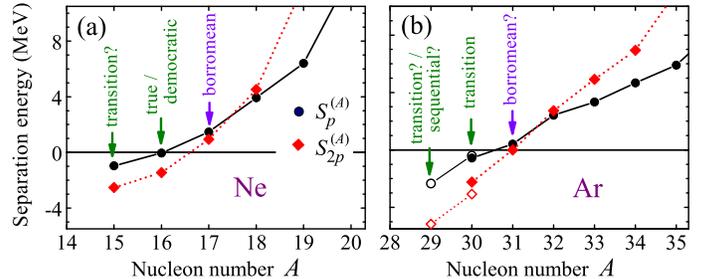}
\end{center}
\caption{(Color online) Systematics of $p$ and $2p$ separation energies ($S^{(A)}_p$ and $S^{(A)}_{2p}$, respectively) in the neon [panel (a)] and argon [panel (b)] isotopes, which are shown by circles and diamonds, respectively. The hollow symbols in panel (b) show the mass predictions from Ref.\ \cite{Tian:2013}.
Isotopes with specific structure and decay properties defined by the ratios of the $S^{(A)}_p$ and $S^{(A)}_{2p}$ values are indicated by the text legends.}
\label{fig:p-2p-syst}
\end{figure}

The ratio  $S^{(A)}_p$/$S^{(A)}_{2p}$ defines different forms of nuclear dynamic in proximity of the proton dripline.
The systematics of the $p$ and $2p$ separation energies for Ne and Ar isotopes near the proton dripline is shown in Fig.\ \ref{fig:p-2p-syst}. These nuclides represent the lower and the upper parts of the $s$-$d$ shell providing a nice illustration of the common trends here. One can see that evolutions of $S^{(A)}_p$ and $S^{(A)}_{2p}$ values are different, and the corresponding curves are intersecting in the proximity of the proton dripline, causing a rapid change of the dynamic properties of nuclides in this region. The closest to the dripline particle-stable nucleus typically has a borromean property (when an extraction of one proton causes the whole three-body system to fell apart) with the condition $S^{(A)}_p>S^{(A)}_{2p}$. Particle-unstable nuclei located just beyond the dripline  often undergo $2p$ decay by the true or democratic mechanism. In addition, a transition situation is possible. Nuclei located  further beyond the dripline pass the transition area and arrive to a domain  of sequential $2p$ decays. The transition dynamic is quite probable situation in the areas between the true (or democratic) and sequential $2p$-decay dominance.

\section{Direct $2p$-decay model}


The derivation of the direct decay model for $2p$ emission was
considered in details in Refs.~\cite{Grigorenko:2007,Grigorenko:2007a}. The model is based on utilization of simplified three-body Hamiltonians, which allows a factorization of the Green's function. Such an approach can be traced back to the ideas of Galitsky and Cheltsov \cite{Galitsky:1964}. The expressions originating from the resonant R-matrix-type approximation \cite{Grigorenko:2007a} look very similar to those arising in the quasi-classical approximation \cite{Fowler:1967}, later used for $2p$ emission in a number of works \cite[and Refs.\ therein]{Brown:2003,Barker:2003}. However, the models may have important differences discussed e.g.\ in \cite{Grigorenko:2007a,Grigorenko:2009b,Pfutzner:2012}.

\begin{figure}[tb]
\begin{center}
\includegraphics[width=0.49\textwidth]{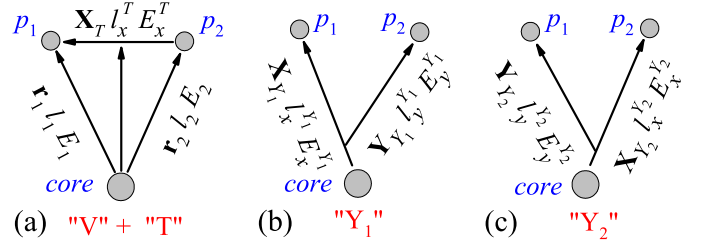}
\end{center}
\caption{(Color online) Three coordinate systems, ``V+T'', ``Y$_1$'', ``Y$_2$'' with respective kinematical variables used in this work for
description of the three-body system core+$p$+$p$ (panels (a), (b), (c), respectively).}
\label{fig:coord}
\end{figure}

The direct decay models of Refs.~\cite{Grigorenko:2007,Grigorenko:2007a} have several important disadvantages which we
overcome in the proposed approach. Namely, the decay amplitudes must be factorized either in the ``V'' system [see Fig.\ \ref{fig:coord} (a)], or in one of the ``Y'' systems [see Fig.~\ref{fig:coord} (b), (c)]. In the ``V'' system, the anti-symmetrization between identical protons is very straightforward. However, the assumption of the infinite core mass is required in such a case. Also the interaction between protons (including Coulomb repulsion) is completely neglected, which leads to some over-estimation of the widths. This type of approximation was called ``no $p$-$p$ Coulomb'' (NPP), see Eq.\ (21) of \cite{Grigorenko:2007}. In the ``Y'' systems the Coulomb problem is effectively resolved because the $Z_iZ_j$ term of Coulomb interaction in $X$ dimension is $Z_{\text{c}}$ while in $Y$ coordinate it is $Z_{\text{c}}+1$. However, the consistent anti-symmetrization is complicated here, and one of the core-$p$ resonances is not treated appropriately. That is because of the $X$ coordinate between the core and $p$. In addition, the $Y$ is only approximate distance between the core and $p$ and coincide with it only in the limit $A_{\text{c}} \rightarrow \infty$ (in this limit, ``Y''$\, \rightarrow \,$``V''). This approximation was entitled ``effective $p$-$p$ Coulomb'' (EFC), see Eq.\ (22) of \cite{Grigorenko:2007}.

In this work we develop the improved direct decay model (IDDM) which is based on simple R-matrix-type analytical approximations for amplitudes and combines positive features of NPP and EFC approximations. We start from the general expression derived in \cite{Grigorenko:2007}, however with modification which correctly accounts for angular momentum coupling of $\{j_1,l_1\}$ and $\{j_2,l_2\}$ for the first and the second core-$p$ subsystems.
\begin{equation}
\frac{d\Gamma(E_{3r})}{ d \Omega_{\varkappa} }=
\frac{8E_{3r}}{\pi (2J+1) } \sum_{M_J}
\left \vert \sum_{\gamma} A^{JM_J}_{\gamma}(\varepsilon,\Omega_1, \Omega_2)
\right \vert ^{2} \;,
\label{eq:corr-flux-imp}
\end{equation}
where $J$ is the total angular momentum, $\gamma=\{j_1,l_1,j_2,l_2\}$, $\Omega_{\varkappa} = \{\varepsilon,
\Omega_1, \Omega_2 \}$, where solid angles $\Omega_i$ are related to vectors $\mathbf{r}_i$, see Fig.\ \ref{fig:coord}. The amplitude $A_{J,\gamma}$ is defined as
\begin{eqnarray}
A^{JM_J}_{\gamma}(\Omega_{\varkappa})  =
\frac{\left [ j_1 \otimes j_2 \right]_{JM_J}}{\sqrt{v_{1}v_{2}}}
\int \nolimits_{0}^{R}dr_1^{\prime} \int \nolimits_{0}^{R}
dr_2^{\prime} \; \varphi_{j_1 l_{1}}(k_{1}r_1^{\prime})\
 \nonumber  \\
\times \; \varphi_{j_2 l_{2}}(k_{2}r_2^{\prime}) \, \Delta V(r_1^{\prime},r_2^{\prime}) \, \varphi_{J\gamma} (r_1^{\prime
}, r_2^{\prime})\,,
\label{eq:amp-1}
\end{eqnarray}
where $\varphi_{J\gamma}$ is assumed to be the radial part of the three-body resonant WF in quasi-stationary approximation in $jj$ coupling. The integral in Eq.\ \ref{eq:amp-1} can be evaluated by replacing the continuum WFs with the quasi-stationary WF in proximity of the two-body resonance energies
\begin{equation}
\varphi_{jl}(kr)=\frac{\sqrt{v(E)}}{2} \, A_{jl}(E)
\;\hat{\psi}_{jl}(E_{r},r)\;,
\label{eq:res-fun}
\end{equation}
where $k$, $v$, $E$ are related momentum, velocity, energy in the two-body channel. The pure radial function $\hat{\psi}_{jl}(E_{r},r)$ is the quasi-bound WF: it has a resonant boundary condition at the two-body resonant energy $E_r$ and is normalized to unity in the internal domain as
\[
\hat{\psi}_{jl}(E_{r},r>R) \propto G_{l}(k_{r}r)\;,\quad \int_{0}^{R}
dr\;\left\vert \hat{\psi}_{jl}(E_{r},r)\right\vert ^{2}=1\;,
\]
where $G_{l}$ is irregular at the origin Coulomb WF with angular momentum $l$.
The amplitude $A_{jl}$ of the resonance with the parameters $E_r$ and $\Gamma_r=\Gamma(E_r)$ is defined as
\begin{equation}
A_{jl}(E) =  \frac{\sqrt{\Gamma_r(E)}} {E_{r}-E-i \Gamma_r(E)/2} +
A^{(p)}_{jl}(E) \,.
\label{eq:res-amp}
\end{equation}
The width as a function of energy is defined by the standard R-matrix expression
\begin{equation}
\Gamma_r(E) = 2 \, \frac{\theta^2}{2Mr^2_{\text{cp}}} \, P_l(Z_1,Z_2,r_{\text{cp}},E)\,,
\label{eq:r-matr-gamma}
\end{equation}
via penetrability function $P_l$, channel radius $r_{\text{cp}}$, and reduced width $\theta$.
In Eq.\ (\ref{eq:res-amp}) the resonant term typically used in such calculations is augmented by taking into account the ``potential scattering''
contribution $A^{(p)}_{jl}$. This allows additional tests of stability of the obtained results. The potential scattering term can be reasonably approximated by the scattering on solid sphere with radius $r_{\text{sp}}$:
\begin{equation}
A^{(p)}_{jl}(E) = - \frac{2F_l(kr_{\text{sp}})}{\sqrt{\Gamma_r(E)}} \,
\frac{G_l(kr_{\text{sp}}) - iF_l(kr_{\text{sp}})}
{F^2_l(kr_{\text{sp}}) + G^2_l(kr_{\text{sp}})}\,.
\label{eq:tot-amp}
\end{equation}

Using Eq.\ (\ref{eq:res-fun}) the amplitude is factorized into momentum-dependent term and radial integral
\begin{eqnarray}
A^{JM_J}_{\gamma}(\Omega_{\varkappa})  =  \frac{V^{J}_{\gamma}}{4} \, \left [ j_1 \otimes j_2 \right]_{JM_J} A_{j_1l_1}(E_1)\,
A_{j_2l_2}(E_2),  \nonumber  \\
V^{J}_{\gamma} =
\int \nolimits_{0}^{R} dr_1^{\prime} \int \nolimits_{0}^{R}
dr_2^{\prime} \;
\hat{\psi}_{j_1l_{1}}(E_{r1},r_1^{\prime}) \, \hat{\psi}_{j_2l_{2}}(E_{r2},r_2^{\prime}) \nonumber  \\
\times \, \Delta V(r_1^{\prime},r_2^{\prime})\;\varphi_{J \gamma}(r_1^{\prime},r_2^{\prime})\,.
\quad
\end{eqnarray}
%

Because the spin degrees of freedom can not be observed even in the most sophisticated modern radioactive ion beam experiments, we may perform the summation over the final state spin quantum numbers. For systems with zero core spin there are two types of terms to be treated.
The first type is with $S=0$ where $J=L$, and the second is with $S=1$ where $L=\{J-1,J,J+1\}$. After averaging over $M_J$, we get
\begin{eqnarray}
\frac{d\Gamma(E_{3r})}{d \Omega_{\varkappa}} = \sum_{LS}
\frac{E_{3r}}{2\pi(2L+1)} \sum_{M_L}
\left \vert \sum \nolimits _{\gamma} A^{LM_L}_{S\gamma}(\Omega_{\varkappa})
\right \vert ^{2}, \quad
\label{eq:corr-flux-imp-1} \\
A^{LM_L}_{S\gamma}(\Omega_{\varkappa}) = C^{JLS}_{\gamma} \,  V^{J}_{\gamma}
\left [ l_1 \otimes l_2 \right]_{LM_L} A_{j_1l_1}(E_1) A_{j_2l_2}(E_2) \,,
\label{eq:corr-flux-imp-2} \\
\left [ l_1 \otimes l_2 \right]_{LM_L} =  \sum \nolimits _{m_1 m_2} C^{LM_L}_{l_1 m_1 l_2 m_2} Y_{l_1 m_1}(\hat{r}_1)Y_{l_1 m_1}(\hat{r}_1)\,.
\nonumber
\end{eqnarray}
The coefficient $C^{JLS}_{\gamma}$ provides the $jj$ to $LS$ re-coupling
for two protons in the orbitals $j_1$ and $j_2$:
\[
C^{JLS}_{j_1l_1j_2l_2} = \hat{L} \hat{S} \hat{j}_1 \hat{j}_2\left \{
\begin{array}{ccc}
l_1  & l_2 & L \\
1/2  & 1/2 & S \\
j_1  & j_2 & J
\end{array}
\right \}\, .
\]
The potential matrix elements $V_{\gamma}^{J}$ can be reasonably approximated \cite{Grigorenko:2007,Brown:2015a} as %
\[
V_{\gamma}^{J} = c_{\gamma}^{J} \sqrt{[E_{r1}+E_{r2}-E_T]^2 + [\Gamma_{1}(E_{r1})+\Gamma_{2}(E_{r2})]^2/4} \,,
\]
where $c_{\gamma}^{J}$ are phenomenological complex coefficients, all normalized to unity $\textstyle \sum _{\gamma} |c_{\gamma}^{J}|^2 \equiv 1$.
The Eq.\ (\ref{eq:corr-flux-imp-2}) uses intermediate states in the subsystems with appropriate sets of quantum numbers and correctly couples them to the total spin of the system. However, there are important drawbacks in the approximation which do not allow to reproduce realistic momentum distributions of fragments in $2p$ decays.

We propose a phenomenological improvement of Eq.\ (\ref{eq:corr-flux-imp-2}) by
replacing the amplitude with the more realistic one, composed with terms not in ``V'' system, but in different Jacoby systems. For the case without potential scattering contribution in Eq. (\ref{eq:res-amp}) it reads
\begin{eqnarray}
A^{LM_L}_{S\gamma}(\Omega_{\varkappa})
\rightarrow \frac{C^{JLS}_{\gamma} \,  V^{J}_{\gamma} \, A_S^{(pp)}(E^T_x)}{
E_{r1}+E_{r2}-E_T - i\,[\Gamma_{1}(E_{r1})+\Gamma_{2}(E_{r2})]/2} \nonumber \\
\times \, \mathcal{\hat{O}}_S \, \left( \left [ l^{Y_1}_x \otimes l^{Y_1}_y
\right]_{LM_L}
A_{j^{Y_1}_x l^{Y_1}_x}(E^{Y_1}_x)\sqrt{\Gamma_{1}(E^{Y_1}_y)} \right. \nonumber  \\
 + \left. \left [ l^{Y_2}_x \otimes
l^{Y_2}_y \right]_{LM_L}  A_{j^{Y_2}_yl^{Y_2}_y}(E^{Y_2}_x) \sqrt{\Gamma_{2}(E^{Y_2}_y)}
\right) \,. \qquad
\label{eq:amp-rewrite}
\end{eqnarray}

The permutation operator $\mathcal{O}_S$ provides a correct symmetry of WF with
respect to proton permutations. As the isospin WF of two
protons are symmetric, we should use
\[
\mathcal{\hat{O}}_0 \equiv \mathcal{S} \;, \qquad \mathcal{\hat{O}}_1 \equiv \mathcal{A}
\]

The amplitude $A^{(pp)}_{S}$  introduces corrections for the $p$-$p$ final state
interactions both for the $S=0$ and $S=1$ states of two protons. The following expression is used:
\begin{equation}
A^{(pp)}_{S}(E^{T}_{x})  =  \frac{N}{v^{T\,}_{x\,}}\sqrt{\frac{2}{\pi}} \int_0^{\infty} dr\, \psi^{(pp)}_{l=S}(k^{T\,}_{x\,}r) \, \phi(r)  \,,
\label{eq:amp-pp}
\end{equation}
where $k^{T}_{x}$ and $v^{T}_{x}$ are momentum and velocity associated with $p$-$p$ energy  $E^{T}_{x}$. The meaning of such an approximation is to consider protons as being emitted from a broad spatial region defined be the function  $\phi(r)$. The source function $\phi(r)$ with Gaussian type formfactor is used
\begin{equation}
\phi(r)  =  \frac{1}{a_{pp}^{3/2}} \sqrt{\frac{54}{\pi}} \, r \, \exp \left(-\frac{3r^2}{4a_{pp}^2} \right)\,.
\label{eq:source-pp}
\end{equation}
The radius parameter $a_{pp}$ in the above expression is equal to the root mean square radius for the function $\phi(r)$. The simple singlet $p$-$p$ potential $V(r)=-V_0 \, \exp[-(r/r_0)^2]$ with $V_0=-31$ MeV and
$r_0=1.8$ fm is used to define the $s$-wave proton scattering WF $\psi^{(pp)}_{l=0}$. The zero nuclear interaction is assumed for the $p$-wave $p$-$p$ motion $\psi^{(pp)}_{l=1}$.
The normalization coefficient $N$ is chosen in such a way that for fixed three-body decay energy $E_T$
\begin{equation}
\int_0^{1} d \varepsilon \, \sqrt{\varepsilon(1-\varepsilon)} \,| A^{(pp)}_{S}(\varepsilon E_T ) |^2 = \pi/8   \,.
\label{eq:amp-pp}
\end{equation}
In the calculations we use $a_{pp}=4$ fm for which $| A^{(pp)}_{S}(E) |^2$ resemble very much the free $p$-$p$ scattering cross section profile; the difference is at small $E$ values, where $A^{(pp)}_{S}(E)$ has correct asymptotic behavior.

The proposed IDDM unifies the advantages of both the NPP and EFC models and also goes beyond, providing a reliable tool for estimates and phenomenology of $2p$ decays:

\noindent \textbf{(i)} It can be shown that Eq.\ (\ref{eq:amp-rewrite}) returns
exactly the (\ref{eq:corr-flux-imp-2}) in the approximation $A_{\text{c}} \gg 1$, $Z_{\text{c}} \gg 1$, $A^{(pp)}_S \rightarrow 1$. Thus the improved expression has clear transition conditions to the previously used models.

\noindent \textbf{(ii)} Spins of the resonances in the subsystems and total angular momentum coupling schemes are correctly accounted in the proposed expressions as well as the proper symmetries of two-proton amplitudes.

\noindent \textbf{(iii)} In the proposed model two resonances in two core-$p$ subsystems can be treated simultaneously without need to use the infinite core mass assumption.

\noindent \textbf{(iv)} The proton-proton final state interaction is treated in phenomenologically sufficient way.

\begin{figure}
\begin{center}
\includegraphics[width=0.48\textwidth]{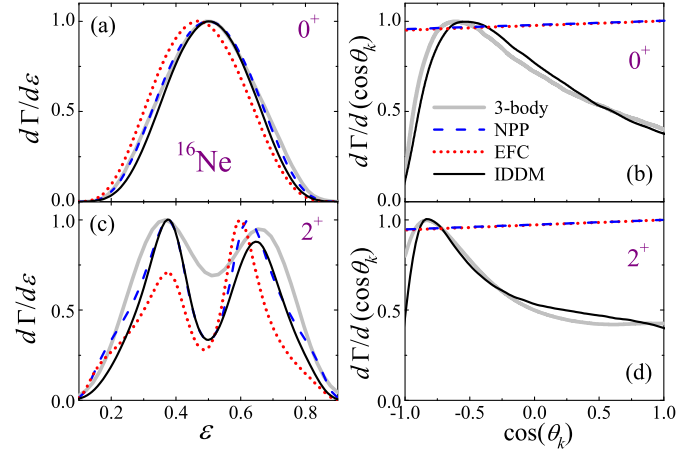}
\end{center}
\caption{(Color online) Energy and angular distributions in the Jacobi ``Y'' system for decay of the $^{16}$Ne $0^+$ g.s.\ and $2^+$ first excited state calculated by the 3-body, NPP, EFC, and IDDM models.}
\label{fig:16ne-all-distr}
\end{figure}


\section{The $^{16}$Ne decays}


Properties of the IDDM are demonstrated by the example of $^{16}$Ne $0^+$ g.s.\ and $2^+$ first excited state, both decaying into the $^{14}$O+$p$+$p$ channel. The $^{16}$Ne decay data have been compared  with the sophisticated full-scale three-body calculations in Refs.\ \cite{Brown:2014,Brown:2015a}. The obtained agreement was quantitative, so in this work we just compare the semi-analytical model results with the complete three-body calculations in order to avoid the Monte Carlo (MC) simulation procedure needed for taking into account the response of the experimental setup. The comparison of energy and angular distributions in the ``Y'' Jacobi system is presented in Fig.\ \ref{fig:16ne-all-distr} and Table \ref{tab:16ne}. In both cases we assume decay via single configuration: $[s^2_{1/2}]_0$ for $0^+$ state and $[s_{1/2}d_{5/2}]_2$ for $2^+$ state. The corresponding $\{E_r,\Gamma_r\}$ sets are $\{1.405,0.7\}$ for $s_{1/2}$ and $\{2.8,0.37\}$ for $d_{5/2}$ in $^{15}$F. The weights of the $[s^2_{1/2}]_0$ and $[s_{1/2}d_{5/2}]_2$ configurations are taken as $67\%$ \cite{Grigorenko:2015} and $16\%$ \cite{Brown:2015a}.

We point to the following aspects of the compared models, see Table \ref{tab:16ne}. The NPP model has a trend to overestimate the decay width \cite{Grigorenko:2007a}. The results of the proposed IDDM residue in between the NPP and EFC width values. NPP and EFC models has different deficiencies in describing the momentum distributions in the $2p$ decay. The IDDM model reproduces all qualitative features of the momentum distributions correctly. E.g.\ pay attention to correct positions and ratio of two peaks in Fig.\ \ref{fig:16ne-all-distr} (c). For $2^+$ state the quantitative agreement with full three-body calculations is not expected as it is understood that several quantum configurations are needed for description of the data \cite{Brown:2015a}.

So, we can see by the example of $^{16}$Ne decay that the proposed IDDM combines positive features of the NPP and EFC models, and it provides the results close to the sophisticated three-body calculations. This makes the proposed approach to be a reasonable substitute for the complicated three-body calculations in the case of exploratory and systematics studies.

\begin{table}[b]
\caption{Widths of $^{16}$Ne (in keV) states calculated with the NPP, EFC, IDDM and 3-body models. The decay energy of the $0^+$ state is $E_T=1.466$ MeV, and the $2^+$ state is $E_T=3.16$ MeV.}
\medskip
\begin{center}
\begin{tabular}[c]{cccccc}
\hline
\hline
$\Gamma$ (keV)  & NPP   & EFC      & IDDM         & 3-body & Exp. \\
\hline
$0^+$           & 4.6   &   2.4    &     4.1     &  3.1 & $<80$ \cite{Brown:2014} \\
$2^+$           & 14    &   11     &     12      &  56  & $ 175 \pm 75$ \cite{Brown:2015a}\\
\hline
\hline
\end{tabular}
\end{center}
\label{tab:16ne}
\end{table}


\section{Transition patterns in the $s$-$d$ shell nuclei}


\begin{figure*}
\begin{center}
\includegraphics[width=0.235\textwidth]{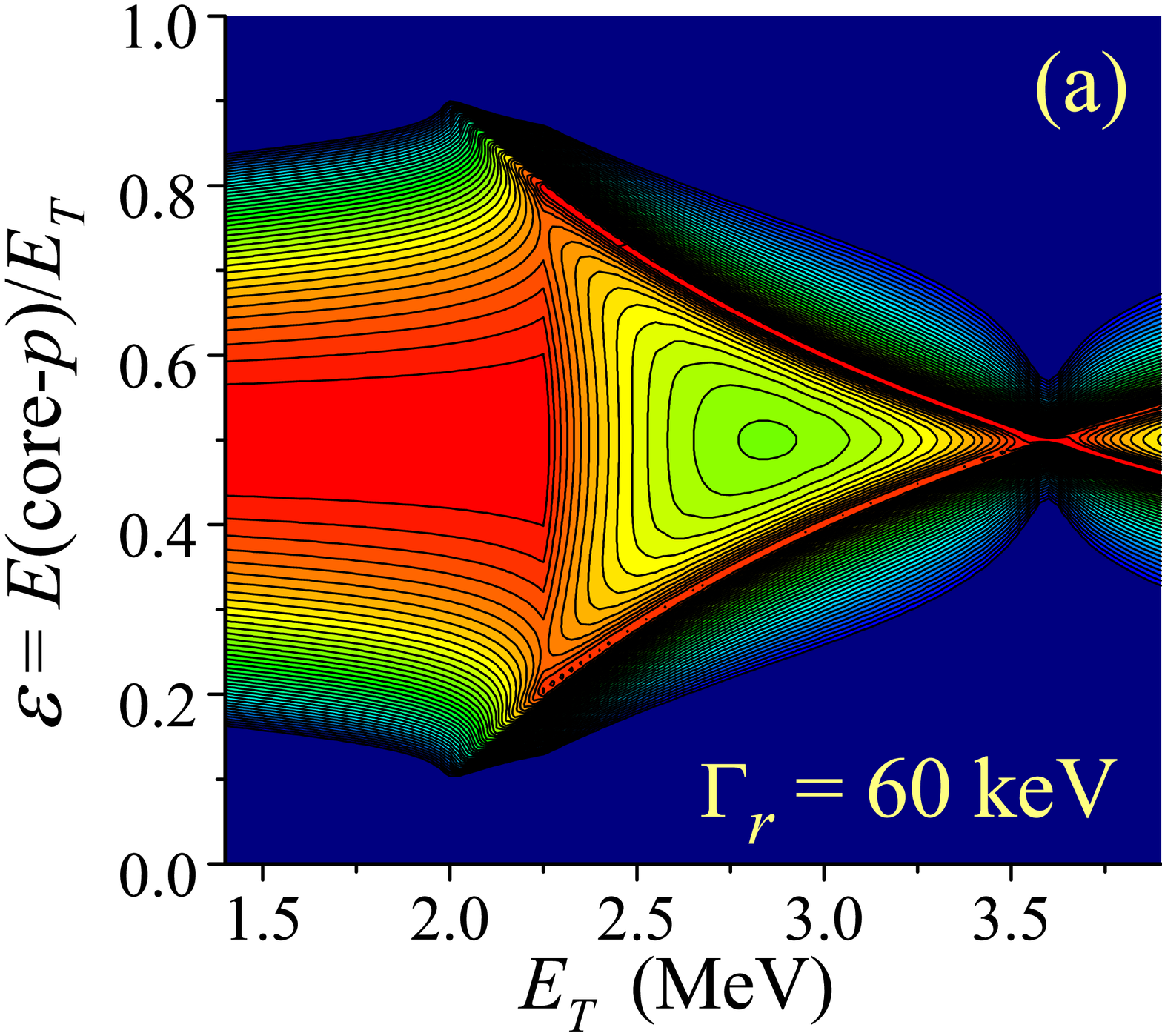}
\includegraphics[width=0.205\textwidth]{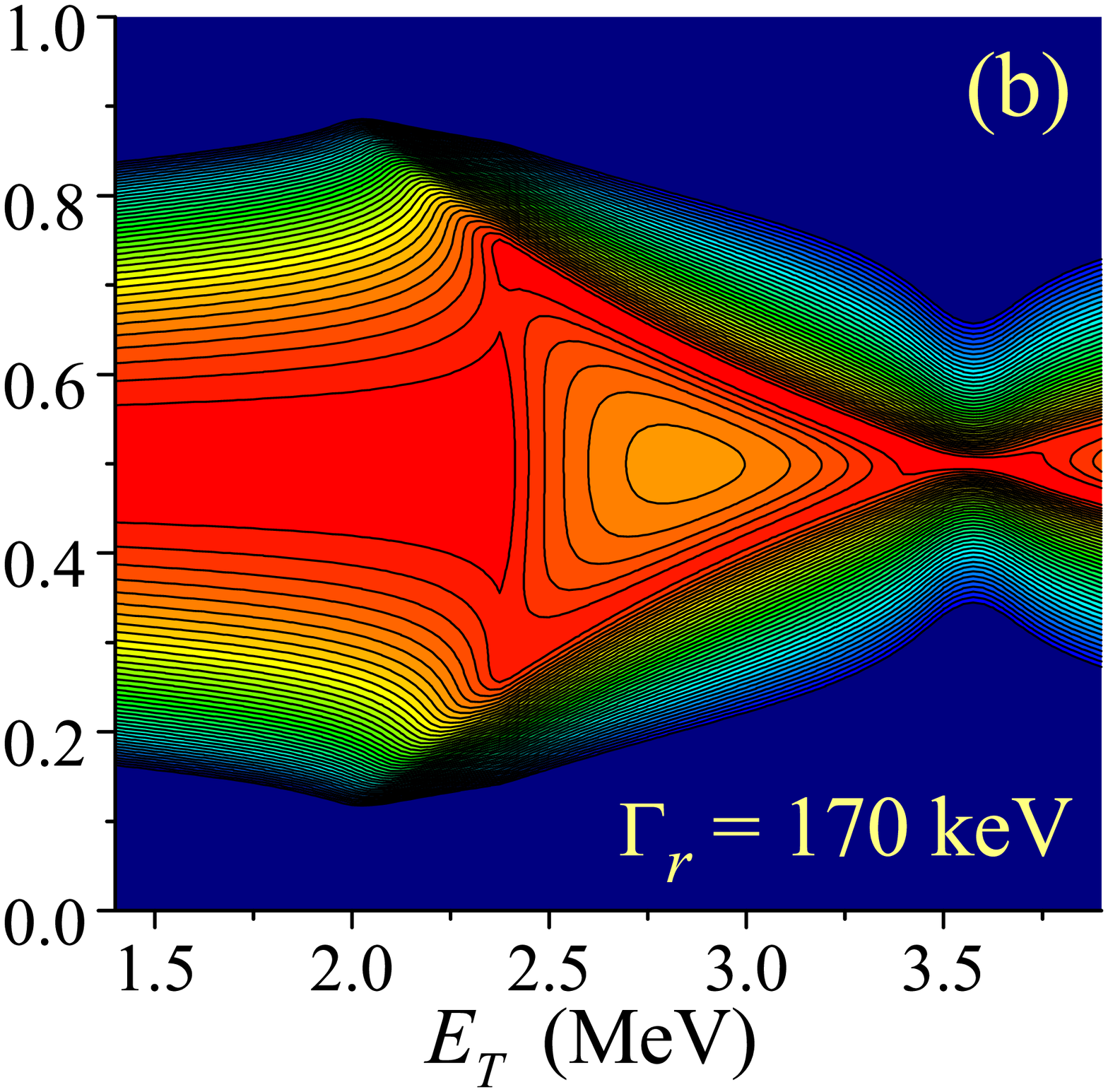}
\includegraphics[width=0.205\textwidth]{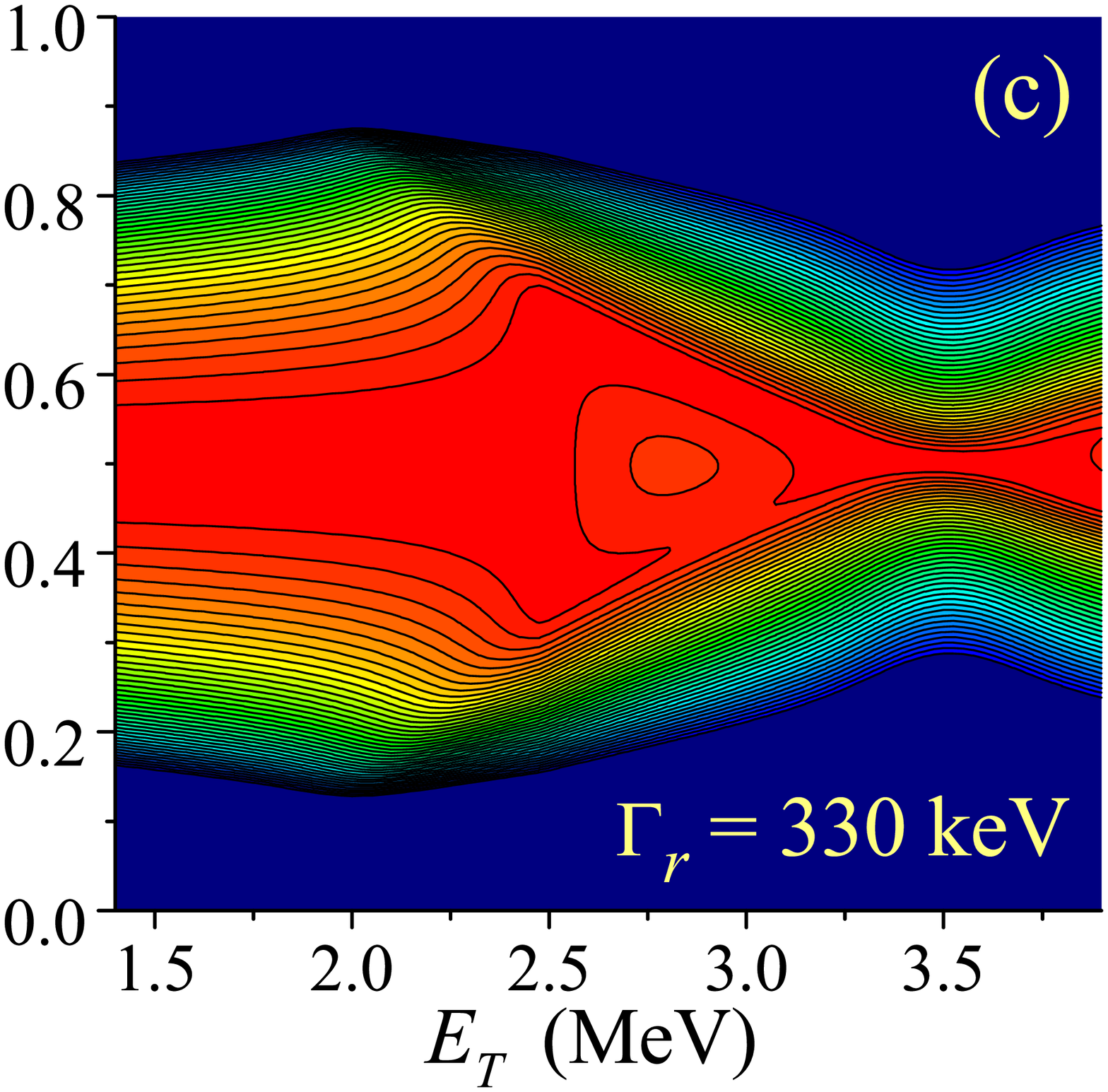}
\includegraphics[width=0.28\textwidth]{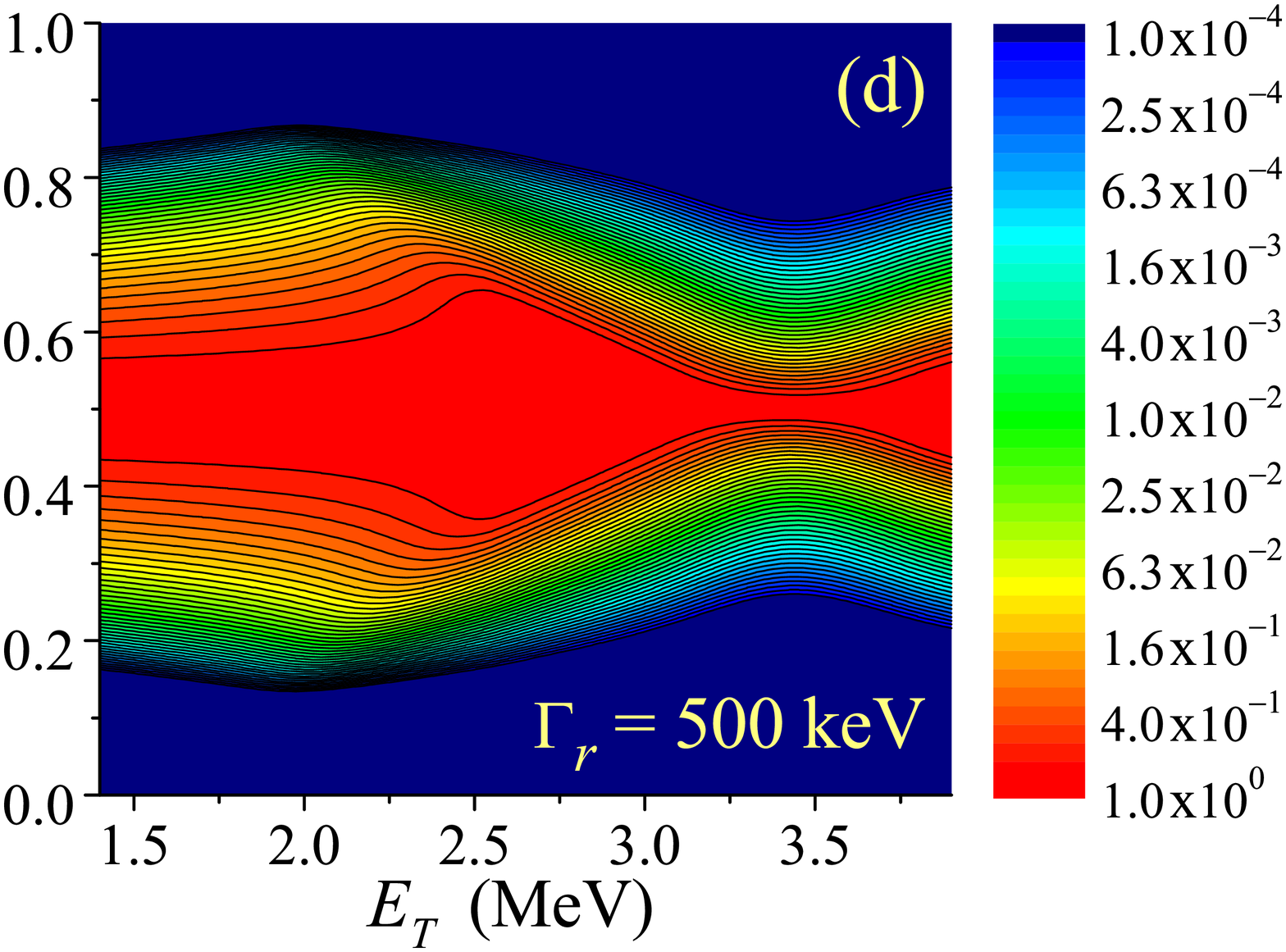}
\end{center}
\caption{(Color online) Energy correlations as functions of the $2p$-decay energy $E_T$ calculated  by IDDM in the ``Y'' Jacobi system for $^{30}$Ar. The $^{29}$Cl g.s.\ energy is $E_r=1.8$ MeV. The panels (a, b, c, d) correspond to the assumed $\Gamma_r$($^{29}$Cl) values of 60, 170, 330, 500 keV, respectively.}
\label{fig:cor-syst-30ar}
\end{figure*}

The transitions between different decay mechanisms are illustrated in Fig.\ \ref{fig:cor-syst-30ar} by calculated energy correlations of fragments of the $^{30}$Ar decaying into $^{28}$S+$p$+$p$. For the fixed energy $E_r=1.8$ MeV, we provide series of plots in the $\{E_T,\varepsilon \}$ dimensions.
The following evolution pattern can be seen in Fig.\ \ref{fig:cor-syst-30ar}(a). At low $E_T$ values, the decay mechanism is ``true $2p$'', which is characterized by a relatively narrow  bell-shaped peak centered at $\varepsilon \sim 1/2$ (when energies of protons are equal). For $E_T>E_r$, two new peaks connected with a sequential emission of protons arise on the side slopes of the central bell-shaped bump. At energy $E_T \sim 1.2 E_r = 2.2 $ MeV, the sequential-emission peaks become higher than the peak at $\varepsilon \sim 1/2$, and this energy value marks a transition point to the sequential decay mechanism (see also the sketch ``True $2p$''$\leftrightarrow$``Sequential $2p$'' transition in Fig.\ \ref{fig:phase-diagram}). In the correlation plot in Fig.\ \ref{fig:cor-syst-30ar}(a), the triangular structure can be seen, which corresponds to  evolution of the double-peak sequential decay pattern as function of $E_T$. At the energy $E_T=2E_r$, the two sequential-emission  peaks overlap, then split again at higher energies $E_T$.

\begin{figure*}
\begin{center}
\includegraphics[width=0.235\textwidth]{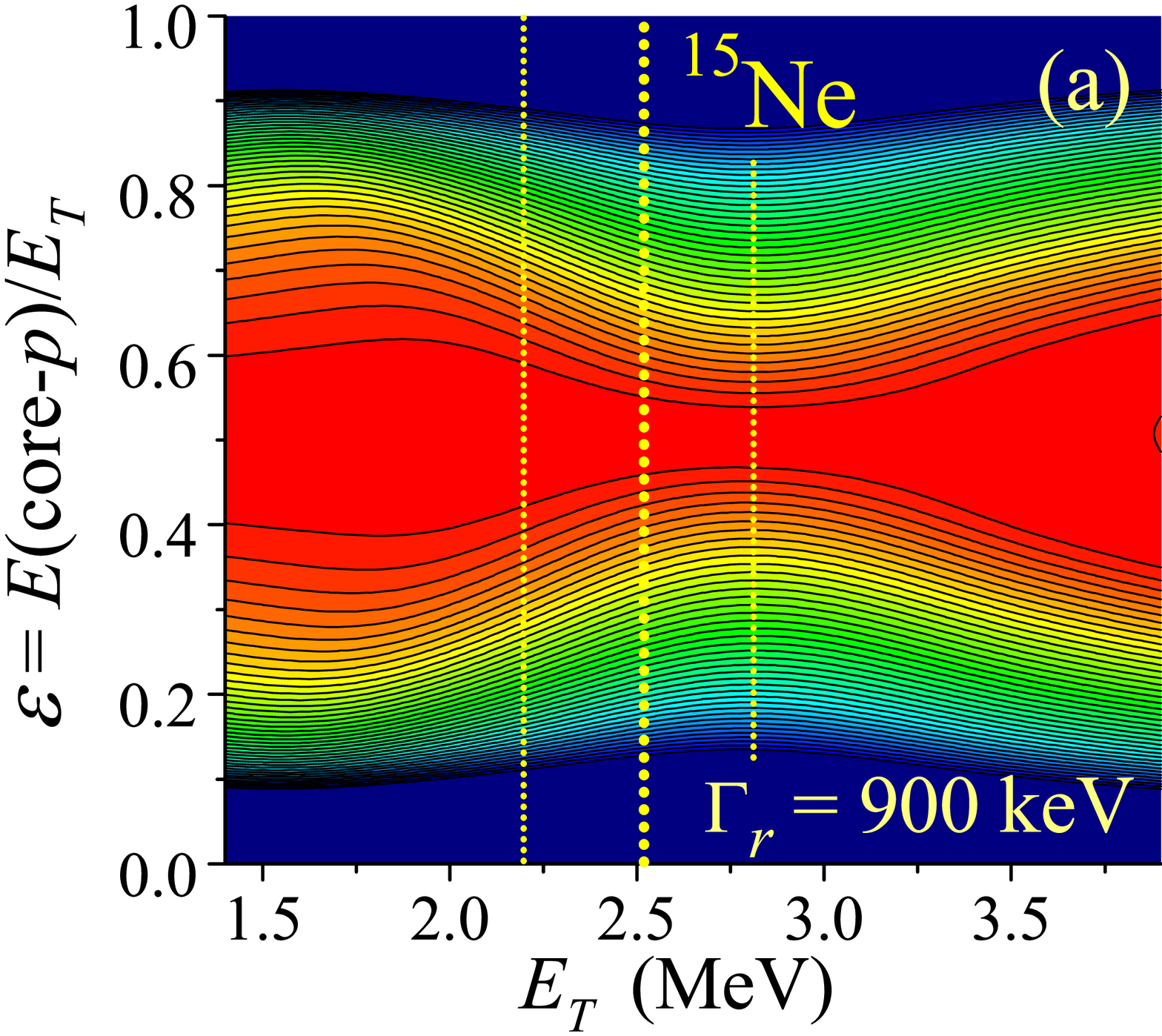}
\includegraphics[width=0.205\textwidth]{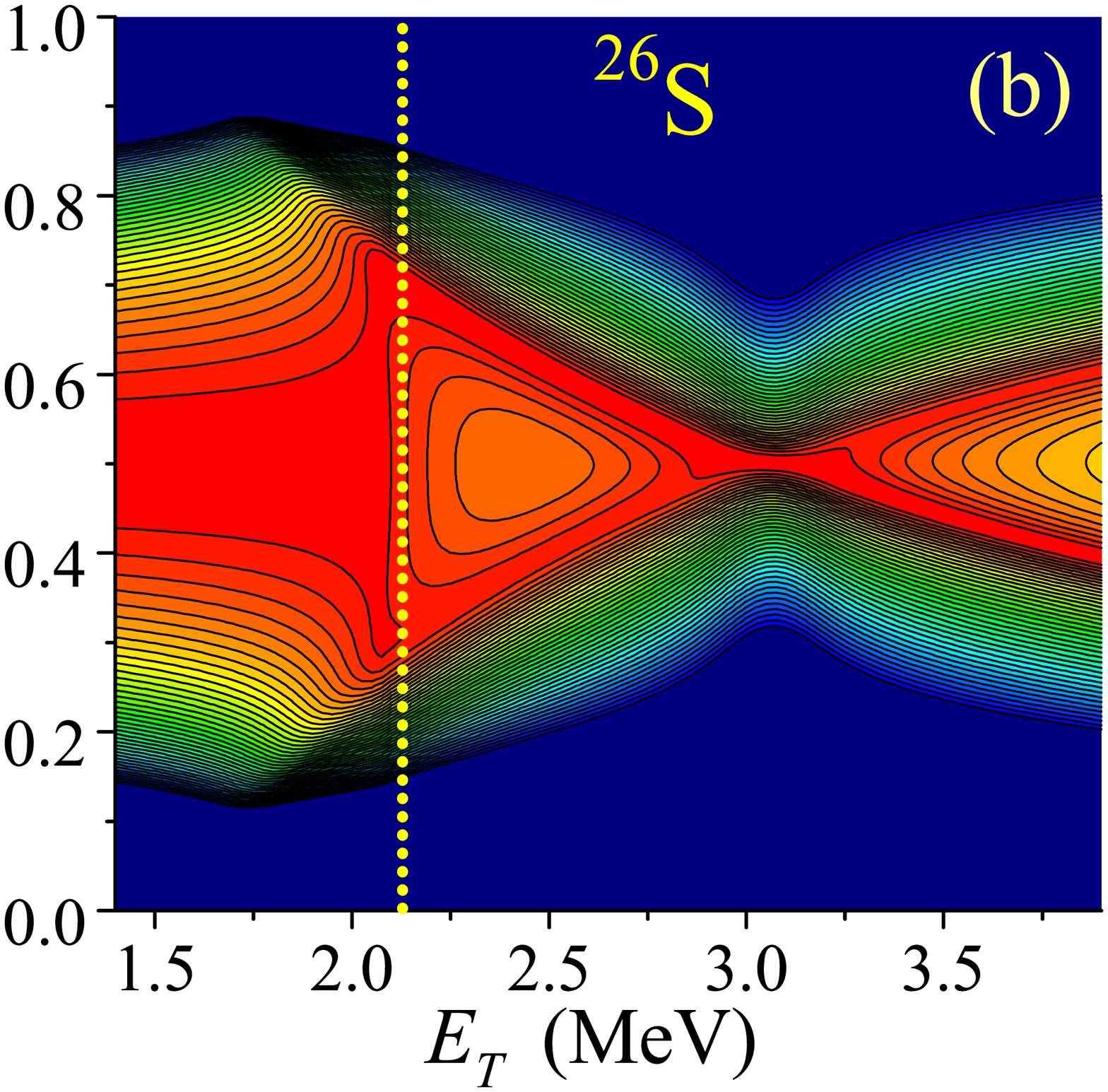}
\includegraphics[width=0.28\textwidth]{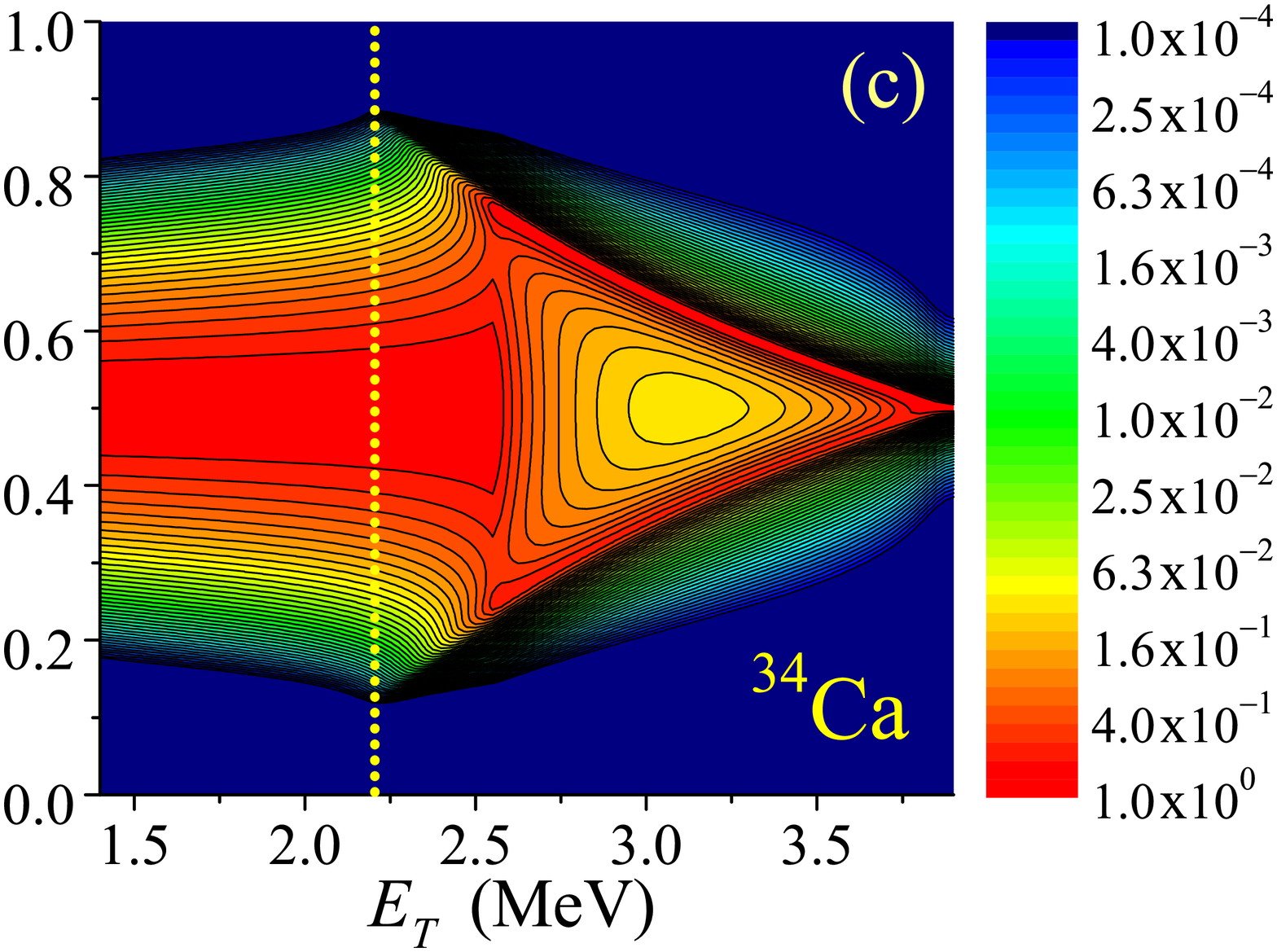}
\end{center}
\caption{(Color online) Energy correlations as functions of the $2p$-decay energy $E_T$ calculated  by IDDM in the ``Y'' Jacobi system for $^{15}$Ne, $^{26}$S, and $^{34}$Ca. The used $\{E_r, \Gamma_r\}$ values are $\{1.56, 0.9 \}$  \cite{Goldberg:2010} (experiment),  $\{1.96, 0.12 \}$ \cite{Wang:2012} (theory), $\{1.55, 0.19 \}$ \cite{Cole:1997} (theory), respectively. The vertical dotted lines mark the $E_T$ values accepted from the experiment on $^{15}$Ne \cite{Wamers:2014} as well as from the systematics evaluations of $^{26}$S \cite{Tian:2013} and $^{34}$Ca \cite{Wang:2012} $2p$ decays. Thin dotted lines for $^{15}$Ne indicate the experimental value of the g.s.\ width $\Gamma=0.59$ MeV \cite{Wamers:2014}. }
\label{fig:cor-syst-all}
\end{figure*}

Figs.\ \ref{fig:cor-syst-30ar} (b, c, d) corresponds to the same parameters, except for the larger widths $\Gamma_r$, which is achieved by variation of the reduced width  $\theta^2$ parameter in Eq.\ (\ref{eq:r-matr-gamma}). We can see that with increase of $\Gamma_r$, the triangular sequential-decay pattern arises at higher energy $E_T$ but gradually smears out, and finally it vanishes. This qualitative change means the transition to the democratic decay mechanism. The evolution observed in the decay patterns from (b) to (d) corresponds to the ``Sequential $2p$''$\leftrightarrow$``Democratic $2p$''  transition region sketched in Fig.\ \ref{fig:phase-diagram}.

The panels (a) and (b) in Fig.\ \ref{fig:cor-syst-30ar} correspond to the $\theta^2$ values of 0.5 and 2.0, respectively. The first value should correspond to nuclear structure with strong configuration mixing. The latter value is the upper limit of the admissible range for $\theta^2$, which corresponds to the pure single-particle structure. The difference between the discussed $E_T$ evolution patterns is quite large. E.g., the transition ``True $2p$''$\leftrightarrow$``Sequential $2p$'' in the calculation (b) takes place at $E_T$ value about 300 keV larger than in the case (a). The unrealistically large  $\theta^2$ values of 4 and 6 are used in the cases (c) and (d). They are provided for illustration purpose of the transition to the democratic $2p$-decay mechanism. It is obvious that the charge of $^{30}$Ar is too large to allow the democratic decay mechanism be possible with the used $E_r=1.8$ MeV value.

 The evolution of $2p$-decay mechanisms similar to the one shown in Figs.\ \ref{fig:cor-syst-30ar} (a, b, c, d) can be achieved by variation of a    charge of $2p$ precursor. Fig.\ \ref{fig:cor-syst-all} illustrates this issue by presenting several examples of the calculated transition patterns ranging from light to heavy  $s$-$d$ shell $2p$ emitters. In the calculated examples of $^{15}$Ne, $^{26}$S, and $^{34}$Ca g.s.\ decays, we can see that the observed ($^{15}$Ne) as well as the evaluated $^{26}$S,  $^{34}$Ca decay energies are located near the transition points. This means that all  considered cases are characterized by strong sensitivity of the decay correlation patterns to the parameters of involved nuclear states.


\section{Decay of the $^{15}$Ne ground state}


\begin{figure}
\begin{center}
\includegraphics[width=0.47\textwidth]{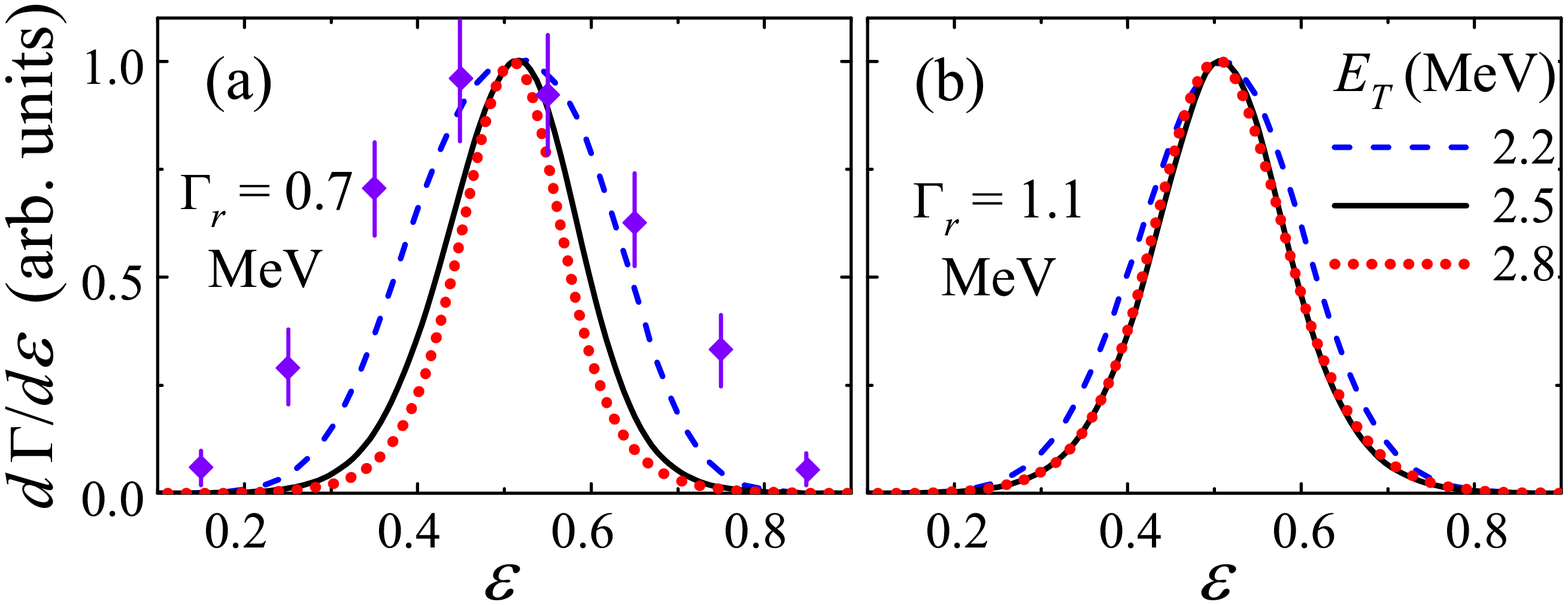}
\end{center}
\caption{(Color online) Energy distributions of fragments in the $2p$ decay of $^{15}$Ne (dots with statistical uncertainties in the ``Y'' Jacobi system \cite{Wamers:2014}) calculated at three $E_T$ values shown in MeV. The $^{14}$F g.s.\ width is assumed either to be 0.7 MeV (a) or 1.1 MeV (b).}
\label{fig:15ne-evol}
\end{figure}

First illustration of a possibility to utilize the $2p$-decay transition dynamic in order to study  a core+$p$ subsystem of the core+$p$+$p$ nucleus is provided by example of the $^{15}$Ne g.s.\  One can see in Fig.\ \ref{fig:cor-syst-all}, that the $2p$-decay energy of $^{15}$Ne g.s., $E_T=2.52$ MeV \cite{Wamers:2014} is located in the region where the $\varepsilon$ distribution strongly depends on $E_T$. The shape of the $\varepsilon$ distribution at fixed energy can be considered as not very reliable model-dependent indicator. However, the $^{15}$Ne g.s.\ is quite broad, $\Gamma=0.59$ MeV, and one may investigate the evolution of the $\varepsilon$ distribution as function of $E_T$ depending on model parameters. Figure \ref{fig:15ne-evol} shows examples of such distributions calculated for  $\Gamma_r(^{14}\text{F}) \pm 0.2$ MeV, which corresponds to lower and upper limits of the existing data about the $^{14}$F g.s.: $E_r(^{14}\text{F})= 1.56(4)$ MeV, $\Gamma_r(^{14}\text{F})= 0.9(2)$ MeV  \cite{Goldberg:2010}. The $\varepsilon$ distributions calculated in the centroid $E_T$ and in the ``wings'' $E_T \pm \Gamma/2$ of the $^{15}$F g.s.\ resonance may differ from each other. They are also quite different in the case of different $^{14}$F g.s.\ widths $\Gamma_r(^{14}\text{F})$, which make this type of information in principle extractable from the decay correlation data.

The experimental distribution for $^{15}$Ne g.s.\ measured in \cite{Wamers:2014} is shown in Fig.\ \ref{fig:15ne-evol}(a). Unfortunately it is strongly broadened by the experimental resolution and the statistics is not sufficient to study the $E_T$ evolution. Evidently this information can not be used for the proposed analysis. However, if we look in the analogous data for the $^{16}$Ne g.s.\ decay \cite{Brown:2014} we can find that much higher accuracy is possible in the modern experiment. The resolution in \cite{Brown:2014} is high enough to potentially resolve the solid and dotted curves in Fig.\ \ref{fig:15ne-evol}(a) or dashed and solid curves in Fig.\ \ref{fig:15ne-evol}(b). So, the proposed type of analysis is principally feasible with modern experimental techniques.


\section{Decay of the $^{30}$Ar ground state}


\begin{figure}
\begin{center}
\includegraphics[width=0.48\textwidth]{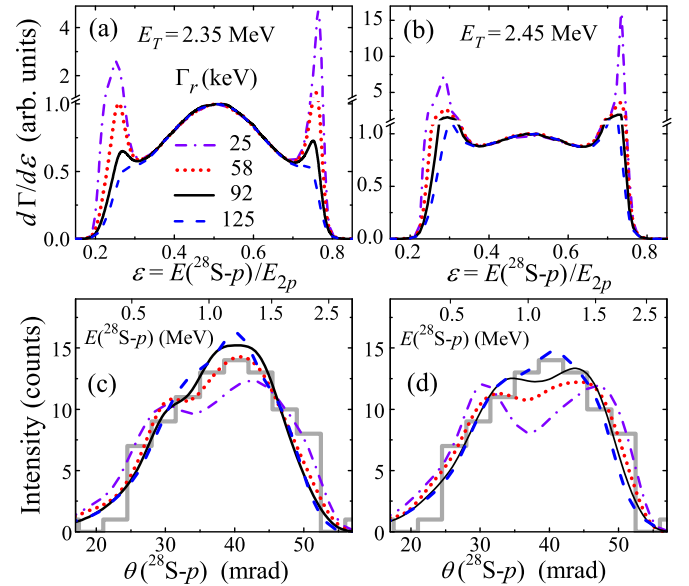}
\end{center}
\caption{(Color online).  Energy distributions of fragments of $2p$ decay of $^{30}$Ar calculated in the ``Y'' Jacobi system for two  $E_T$ values shown in the panels (a) and (b). The $E_r=1.8$ MeV value is fixed, while the curves of different styles correspond to the varied $\Gamma_r$ values. In the panels (c) and (d), the experimental angular distribution measured for $^{30}$Ar g.s.\ in Ref.~\cite{Mukha:2015} is compared with those stemming from respective theoretical distributions in panels (a) and (b) after experimental bias is taken into account via MC simulations.}
\label{fig:gs-distr}
\end{figure}


The observation of $^{30}$Ar and $^{29}$Cl isotopes was reported recently
\cite{Mukha:2015}. It was concluded that the $2p$ decay of the $^{30}$Ar g.s.\ belongs to a transition dynamic. However, the analysis was attempted in a simplified way with the decay energy as the only parameter.
We analyze the $^{30}$Ar data by calculations based on the IDDM. Several calculated correlations and the respective ``experimental'' distributions obtained by simulating the setup response (in the same way as described in \cite{Mukha:2015})  are shown in Fig.\
\ref{fig:gs-distr} together with the experimental data. The experimental setup of
\cite{Mukha:2015} allowed for reconstructing of transverse momentum distributions of
the $2p$-decay products, though the distributions were strongly affected by the setup
response. However, one can see in Fig.\
\ref{fig:gs-distr} that the specific patterns connected with a formation of sequential-decay correlations in the $^{30}$Ar g.s.\ produce the respective changes in the angular distributions $^{28}$S-$p$.

\begin{figure*}
\begin{center}
\includegraphics[width=0.9\textwidth]{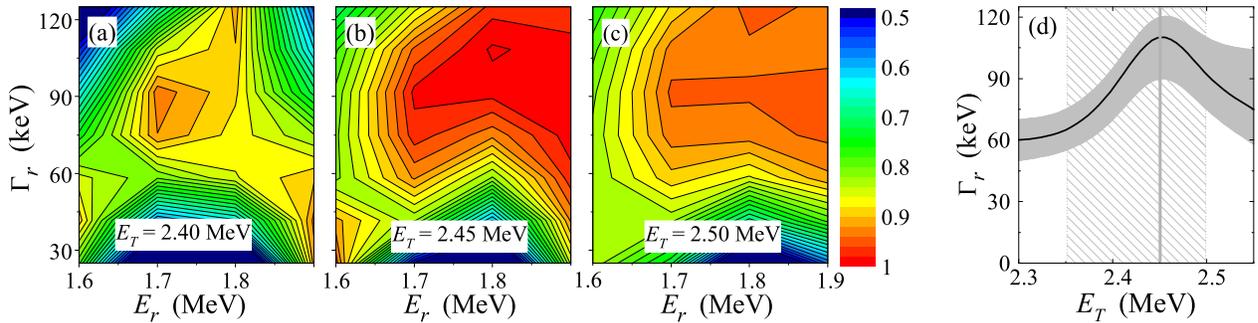}
\end{center}
\caption{(Color online). Panels (a,b,c) show probability of the compatibility of theoretical and experimental (see Fig.\ \ref{fig:gs-distr}) angular distributions from $2p$ decays of the $^{30}$Ar g.s.\  They are functions of  $\{E_r, \Gamma_r\}$ variables for the fixed $E_T$ values indicated in the panels. The panel (d) shows the band of $\{\Gamma_r, \Delta \Gamma_r\}$ values correlated with $E_T$. The best-fit $E_T$ value is shown by the vertical line and its admissible variation by hatching.}
\label{fig:prob-30ar}
\end{figure*}

We have systematically studied reasonable combinations of parameters
$\{E_T,  E_r, \Gamma_r\}$ in order to obtain the best fit of the data. The obtained simulation results have been compared to the data
by the Kolmogorov test, which has provided the probability of the matching simulation and experimental histograms, see Fig.\ \ref{fig:prob-30ar}. The first assigned parameters in Ref.\ \cite{Mukha:2015} were
$E_T=2.25^{+15}_{-10}$ MeV, $E_r=1.8(1)$ MeV. We infer less uncertain value $E_T=2.45^{+5}_{-10}$ MeV  on the basis of  analysis shown in Fig.\ \ref{fig:prob-30ar}, and also the  width of $^{29}$Cl g.s.\ is derived as $\Gamma_r=85(30)$ keV. We emphasize that in this analysis  uncertainties of the major parameters correlate, which can be used for further limitations. For example, it is shown in Fig.\ \ref{fig:prob-30ar}(d) that even smaller $\Delta \Gamma_r \sim 15$ keV can be established for the fixed $E_T$ value.
We note that this type of analysis is in the ``proof of the concept'' stage, and in prospect it should be validated by the direct measurements.


\section{Stability of IDDM results}

\begin{figure}
\begin{center}
\includegraphics[width=0.48\textwidth]{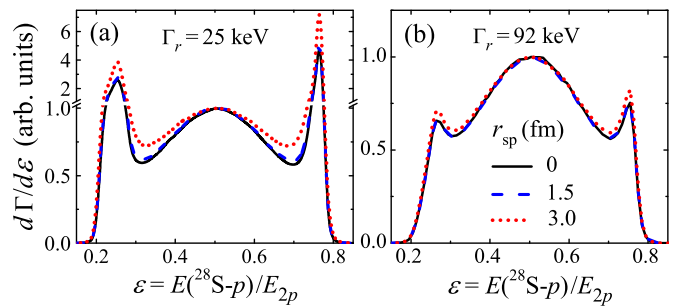}
\end{center}
\caption{(Color online). Sensitivity of energy distributions in the ``Y'' system to modifications of amplitudes (\ref{eq:res-amp}) by ``potential scattering'' contributions (\ref{eq:tot-amp}) with different $r_{\text{sp}}$. Example of $^{30}$Ar (see Fig.\ \ref{fig:gs-distr}) with $E_T=2.35$ MeV, $E_r=1.8$ MeV and two different $\Gamma_r$ values.}
\label{fig:pot-contr}
\end{figure}

The proposed derivation of width of the core-$p$ g.s.\  from the ``Y'' system energy  distributions  depends on the stability of the distributions to the model parameters. We have systematically studied such sensitivity and found that the only significant dependence is to the general decay parameters $\{E_T,E_r,\Gamma_r\}$. E.g., transition patterns strongly depend on the interference of two-body amplitudes at resonant and near-resonant energy. However, Fig.\ \ref{fig:pot-contr} shows that realistic modification of amplitudes outside of the resonance peak does not lead to significant changes in the typical considered situation.

The other important assumption  is that only one quantum configuration provides dominating contribution to the decay width of the state. However, this is a natural situation for the transition dynamic, which is connected to the strong increase of the width just in one selected channel. This issue can be estimated theoretically elsewhere.

\section{Summary}


Near half of the proton-rich $s$-$d$ shell nuclei located by 1--2 atomic numbers beyond the proton dripline  decay by $2p$ emission. Three established mechanisms of nuclear $2p$ decays are the true $2p$, democratic and sequential ones, whose areas of dominance are defined mainly by relations of three general parameters, $E_T$, $E_r$, and $\Gamma_r$. The momentum distributions of the decay fragments in transition regions between these decay mechanisms demonstrate dramatic changes within narrow ranges of the general $2p$-decay parameters.

The behavior of the momentum distributions of $2p$-decay products in the transition regions is described by the improved semi-analytical direct decay formalism. In contrast to the previously-used approximations, the applied model depicts all qualitative features of $2p$-decay distributions in the whole kinematical space. The model works in the mentioned $2p$-decay regimes as well as in their transition regions.

In the transition regions, momentum distributions of the decay products demonstrate strong sensitivity to the variation of general $2p$-decay parameters. Therefore experimental data in transition regions can be used for a precise extraction of resonance parameters in the core-$p$ subsystem of the $2p$-decay precursor. Practical feasibility of such an approach has been demonstrated by the example of the $^{15}$Ne g.s.\ $2p$-decay.
By assuming the transition dynamic, we have established more narrow boundaries for the parameters $E_T$, $E_r$, and $\Gamma_r$ for the $^{30}$Ar and $^{29}$Cl g.s.\ in comparison with those derived in Ref.\ \cite{Mukha:2015}.

%
\textit{Acknowledgments.}
%
%
--- T.A.G.\ was partly supported by the Flerov laboratory JINR and by the HIC for FAIR grant. X.X.\ acknowledge the support by the HIC for FAIR grant. L.V.G.\ was partly supported by thevRussian Ministry of Education and Science grant No.\ NSh-932.2014.2 and by thevRussian Foundation for Basic Research grant No.\ 14-02-00090.



\bibliographystyle{elsart-num-m}
\bibliography{d:/latex/all}


\end{document}